\documentclass[pdflatex,sn-mathphys-num]{sn-jnl}

\usepackage{graphicx}%
\usepackage{multirow}%
\usepackage{amsmath,amssymb,amsfonts}%
\usepackage{amsthm}%
\usepackage{mathrsfs}%
\usepackage[title]{appendix}%
\usepackage{xcolor}%
\usepackage{textcomp}%
\usepackage{manyfoot}%
\usepackage{booktabs}%
\usepackage{algorithm}%
\usepackage{algorithmicx}%
\usepackage{algpseudocode}%
\usepackage{listings}%
\usepackage{multirow}
\usepackage{natbib}
\usepackage{makecell}
\usepackage{xcolor}

\newcolumntype{C}[1]{>{\centering\arraybackslash}m{#1}}

\theoremstyle{thmstyleone}%
\theoremstyle{thmstyletwo}%

\theoremstyle{thmstylethree}%

\begin{document}

\title{Accurate and scalable deep Maxwell solvers using multilevel iterative methods}


\author[1]{\fnm{Chenkai} \sur{Mao}}\email{chenkaim@stanford.edu}

\author*[1]{\fnm{Jonathan A.} \sur{Fan}}\email{jonfan@stanford.edu}

\affil*[1]{\orgdiv{Department of Electrical Engineering}, \orgname{Stanford}, \orgaddress{\street{450 Jane Stanford Way}, \city{Stanford}, \postcode{94305}, \state{CA}, \country{USA}}}


\abstract{
Neural networks have promise as surrogate partial differential equation (PDE) solvers, but it remains a challenge to use these concepts to solve problems with high accuracy and scalability. In this work, we show that neural network surrogates can combine with iterative algorithms to accurately solve PDE problems featuring different scales, resolutions, and boundary conditions.  We develop a subdomain neural operator model that supports arbitrary Robin-type boundary condition inputs, and we show that it can be utilized as a flexible preconditioner to iteratively solve subdomain problems with bounded accuracy. We further show that our subdomain models can facilitate the construction of global coarse spaces to enable accelerated, large scale PDE problem solving based on iterative multilevel domain decomposition. With two-dimensional Maxwell's equations as a model system, we train a single network to simulate large scale problems with different sizes, resolutions, wavelengths, and dielectric media distribution. We further demonstrate the utility of our platform in performing the accurate inverse design of multi-wavelength nanophotonic devices. Our work presents a promising path to building accurate and scalable multi-physics surrogate solvers for large practical problems.

}




\maketitle

\section{Introduction}\label{sec:Introduction}

Artificial neural networks have emerged as powerful tools for accelerating scientific computing, particularly in solving partial differential equations (PDEs) that govern complex physical processes including thermal transport\cite{cai2021physics, wu2019predicting}, structural mechanics\cite{he2024multi, li2024mechanics}, fluid dynamics\cite{font2025deep, li2024synthetic}, and electromagnetics\cite{zhu2025frequency, trivedi2019data}. As specified by the Universal Approximation Theorem\cite{lu2019deeponet}, these networks can serve as \textit{direct} solution approximators and are theoretically capable of outputting arbitrary PDE solutions with an accuracy level determined by the training dataset and model size. Practically, however, it has been observed that neural network surrogate solvers are difficult to scale for reasons beyond model and training dataset size limitations.  Most models are limited to solving toy problems with fixed domain sizes, fixed or implicit boundary conditions, and a narrow distribution of parameters (i.e., constitutive parameters, source types, etc.), which prevents them from generalizing beyond the training settings.  Neural network solvers also become increasingly difficult to train and are less accurate as the problem size and parameter distribution grow. 

To overcome these limitations, there has been interest to investigate how \textit{iterative} PDE solving methods can be incorporated into machine learning workflows to enhance their capabilities and scalability.  Iterative methods are established methods in conventional PDE solving that were developed to efficiently solve large scale problems, and they encompass different concepts.  One class of concepts involves iterative neural network-enhanced algebraic preconditioners, where the idea is to train deep networks that take PDE residuals as inputs and that output the corresponding error corrections.  These errors can then be used directly with basic iterative methods \cite{hsieh2019learning} or couple with Krylov subspace methods such as conjugate gradient or generalized minimal residual (GMRES) methods \cite{kopanivcakova2025deeponet, luo2024neural, trifonov2024learning}. These concepts have been adapted using all major classes of neural networks, including physics informed neural networks (PINNs) \cite{raissi2019physics}, neural operators \cite{li2020fourier, lu2019deeponet,cao2024laplace} and graph neural networks \cite{sanchez2020learning, bryutkin2024hamlet}.  

A second class of concepts pertains to the iterative solving of large scale PDE problems and include multigrid \cite{dong2024pinn, zhang2024blending}, domain decomposition \cite{jagtap2020extended, basir2023generalized, dolean2024multilevel, pmlr-v235-mao24b}, and multilevel Schwarz methods \cite{luz2020learning}. With multigrid methods, global problems are iteratively solved using a hierarchy of spatial resolutions, and neural network models have been utilized to predict low spatial frequency errors \cite{dong2024pinn, zhang2024blending} and to more generally solve PDE problems at different resolutions \cite{lerer2024multigrid}.   With domain decomposition, global problems are subdivided into a set of subdomain problems that are iteratively evaluated until global self-consistency is obtained. Many machine learning efforts have focused on the solving of subdomain problems, where the online training of PINN models and the use of neural operator-based surrogate solvers \cite{pmlr-v235-mao24b} has been demonstrated at the subdomain level. For multilevel domain decomposition methods, many works focus on using neural networks to reduce the number of subdomain eigenvalue problems to be solved \cite{heinlein2021combining, heinlein2019machine, heinlein2023predicting}.

While these approaches show promise in certain capacities, it remains challenging to adapt them to general problems featuring a wide range of boundary conditions and domain sizes.  It is particularly difficult to apply these techniques to hyperbolic PDEs such as Maxwell’s equations, which is the focus here, where high spatial resolution is required and where the system matrix is indefinite and becomes ill-conditioned for large scale heterogeneous global domains. To date, efforts have been made to solve Maxwell's equations with neural networks integrated with iterative preconditioning\cite{trivedi2019data, azulay2022multigrid}, multigrid\cite{cui2024neural, lerer2024multigrid, xiemgcfnn}, and domain decomposition methods\cite{knoke2023domain, knoke2023domain, piao2024domain, pmlr-v235-mao24b}, with the intent to extend the accuracy and scalability of deep solvers.  While these efforts have demonstrated improved capabilities in certain limits, they still display shortcomings in key metrics pertaining to accuracy, scalability, dielectric media contrast, wavelength range, and boundary condition types. For instance, concepts based on iterative preconditioners and multigrid methods have been limited to fixed domain sizes and do not generalize to arbitrary boundary conditions.  Iterative neural domain decomposition methods show promise in adapting to a wide range of global domain dimensions, but the limited accuracy of the surrogate subdomain solvers limits global domain simulation accuracy and scalability.

\begin{figure}[ht!]
\centering
\includegraphics[width=1.0\textwidth]{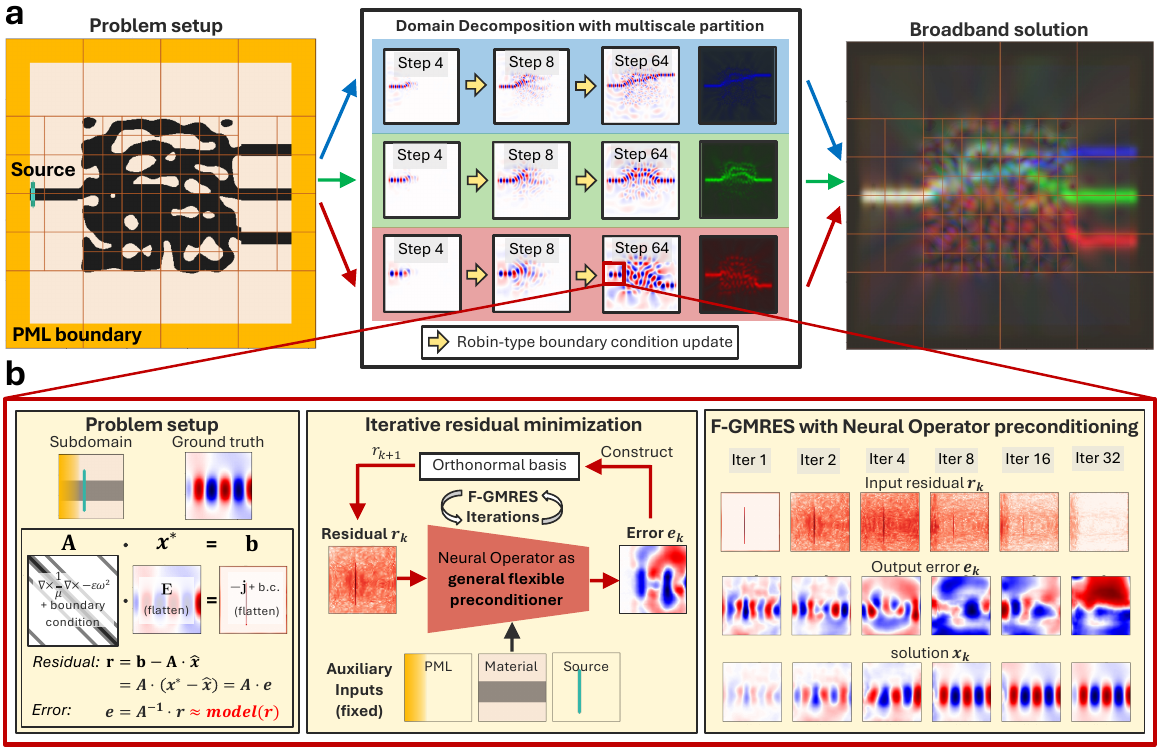}
\caption{PDE solving framework combining a deep subdomain solver with nested inner and outer iterative loops. 
\textbf{a.} Iterative global domain solving using domain decomposition. The global domain is partitioned into overlapping subdomains that are iteratively solved using the overlapping Schwarz method until self-consistency. The subdomain solver also facilitates coarse space construction. Multi-wavelength simulations are achieved by performing multiple global domain simulations at different wavelengths in parallel.
\textbf{b.} Iterative subdomain solving using neural operator-preconditioned F-GMRES. The neural network model uses the residual and auxiliary data as inputs, and it predicts the corresponding error, which serve as Krylov subspace basis directions in F-GMRES. 
}\label{fig1}
\end{figure}

We propose a new approach to neural network-enhanced iterative algorithms that supports the solving of PDE problems with scalability and bounded accuracy.  As a model system, we consider 2D Maxwell's equations for TE polarization (i.e., $E_z$, $H_x$ and $H_y$ field components) in the frequency domain, and we focus on the simulation of dielectric media with subwavelength-scale structuring, which is typical in the fields of nanophotonics and metamaterials. Our concept is outlined in Figure \ref{fig1} and showcases the use of iterative algorithms at both the global domain and subdomain levels to achieve scalable and accurate solving capabilities. At the subdomain level, we train a generalized neural operator that maps residuals to errors and that works for subdomain systems featuring different parameter maps, boundary conditions, resolutions, and wavelengths.  We use the neural network as a general preconditioner in iterative residual minimization schemes to solve subdomain problems with bounded accuracy. At the global domain level, we use domain decomposition with a coarse-space correction to iteratively update subdomain boundary values until the global solution is self-consistent. 

Our approach to iterative PDE solving combines multiple innovations that collectively enable PDE problem solving with unprecedented capabilities.  First, we generalize our neural operator subdomain preconditioner to work for arbitrary Robin-type boundary conditions, which enables the adaptation of these subdomain solvers with advanced methods in domain decomposition such as the overlapping Schwarz method.  We additionally generalize our solver to work for a continuous range of wavelengths, enabling multi-wavelength and broadband evaluation.  Second, we use the trained subdomain networks to efficiently solve subdomain eigenvalue problems with robin-to-robin maps. The solutions to these eigenvalue problems enable the construction of efficient coarse spaces in the global domain, achieving near-optimal scaling in which the iterations required remains constant or grows very slowly with increased problem size. Third, we combine iterative methods at both the subdomain and global domain level, which is a synergistic and critical strategy to global domain scaling because domain decomposition methods require highly accurate subdomain solutions to prevent global-scale error propagation. To demonstrate the capabilities of our platform, we show for the first time the utilization of a deep learning platform that can accurately and efficiently optimize large scale chip-based and free space multi-wavelength nanophotonic devices.

\section{Results}
\subsection{Subdomain neural operators as general preconditioners} \label{sec:subdomain}

We first focus on the solving of subdomain problems, which can each be framed as a boundary value PDE problem with robin-type boundary conditions, $g_i$:
\begin{align}
\nabla\times\nabla\times\mathbf{E}(\mathbf{r}) - \cfrac{\omega^2}{c^2} \varepsilon(\mathbf{r})\mathbf{E}(\mathbf{r}) &= -i\omega\mu_0 \mathbf{J}(\mathbf{r})\label{eq-1}\\
jk(\mathbf{r})\mathbf{E}_i(\mathbf{r}) - \frac{\partial\mathbf{E}_i}{\partial\mathbf{n}} &= g_i\label{eq-2}
\end{align}
Here, $\omega$ is the angular frequency, $\mathbf{E}$ is the electric field, $\varepsilon(\mathbf{r})$ is the heterogeneous relative permittivity, $\mathbf{J}(\mathbf{r})$ is the electric current density, $k(\mathbf{r})=\frac{\omega}{c}\sqrt{\varepsilon(\mathbf{r})}$ is the wavevector on the boundary, and $\mathbf{n}$ is the outward normal vector. Using finite difference frequency domain (FDFD) discretization, the system can be described as a linear equation $Ax=b$, in which $A$ is a sparse matrix that involves the differential operator and boundary operator on the left-hand sides of \eqref{eq-1} and \eqref{eq-2}, and $b$ contains the source term and boundary values from the right-hand sides (Figure \ref{fig1}b, left). 

To solve these problems, our approach is to train a Fourier neural operator that takes auxiliary data (i.e., dielectric distribution, source maps, PML maps) and a residual as inputs and outputs the corresponding error (Figure \ref{fig1}b, center). For a given approximate solution $\hat{x}$, the error can be quantified as $e = x^* - \hat{x}$ and the corresponding residual is $r =A(x^*-\hat{x}) = Ae$ (Figure \ref{fig1}b, left). In this manner, the network approximates the $A^{-1}$ operator on broad residual inputs, serving as a general preconditioner for iterative methods, in our case flexible GMRES (F-GMRES), whereby the error are used to construct basis directions for the Krylov subspace. The algorithmic details are shown in the Methods sections.
As with all preconditioning methods, simulation accuracy can be bounded by evaluating the residual over the course of F-GMRES iterations and stopping the solver once a desired residual threshold has been met (Figure \ref{fig1}b, right).

\begin{figure}[ht!]
\centering
\includegraphics[width=1.0\textwidth]{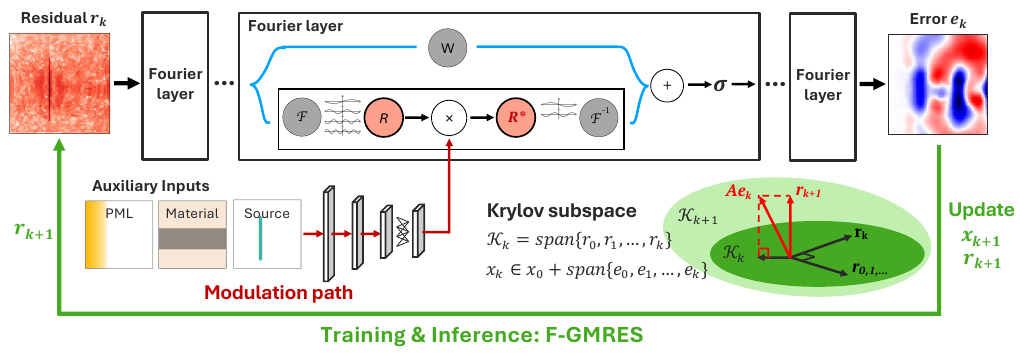}
\caption{The model architecture, which contains a series of Fourier neural operator (FNO) blocks. Auxiliary inputs are processed with convolution, pooling, and fully-connected layers to form a tensor that modulates linearly transformed lower Fourier modes in each FNO block. During training, a dynamic training set of residuals is constructed by iteratively performing F-GMRES on the training data with the model, with the output error vectors being used as Krylov bases in F-GMRES. The same F-GMRES loop is used in inference. $\mathcal{F}$: Fourier transform. $\sigma$: leaky-ReLU
activation function. $\mathbf{W}$: channel mixer via 1-by-1 kernel convolution. $\mathbf{R}$: linear transform on the lower Fourier
modes. $\mathbf{R^*}$: modulated linear transform through an element-wise multiplication with the modulation tensor.}\label{fig2}
\end{figure}

To implement our neural operator, we utilize a modified Fourier neural operator architecture (Figure \ref{fig2}) together with a training scheme that is adapted for photonics problems. Conventionally, the Fourier neural operator architecture utilizes a linear transformation with weight $R$ in the Fourier space, which has learned static weights that apply the same transformation to different input data. In this manner, the size of of $R$ can quickly become a storage bottleneck as the diversity of input data increases and the network scales up. To boost network expressiveness for a diverse input data distribution, we multiply the tensor $R$ with a modulation tensor obtained from a modulation path that applies convolutional, pooling, and fully connected layers to auxiliary inputs. We also adopt a low-rank parameterization of $R$, where a linear layer expands a low-dimensional tensor to construct the full-sized tensor.

To train the network, we construct a training set that is representative of the nanophotonics design space. The training set comprises 1M subdomains and their associated fields, and they include a wide range of freeform dielectric structures, fields, sources, and Robin-type boundary conditions. The dielectric structures consist of discrete and continuous dielectric distributions with permittivity ranging from 1 to 8, and they are obtained by cropping subdomains from large scale FDFD simulations. To perform training, each training data sample is recurrently processed in a F-GMRES loop preconditioned with our model and gradients are backpropagated to minimize both data loss and physical residual loss. Importantly, network outputs are subsequently used to build a Krylov subspace via F-GMRES, yielding new residuals that are recurrently fed back into the network training process. We define the number of total F-GMRES iterations used in the training process as the ``training data recurrence". For the ease of implementation, we found out that a simple fixed point iteration training scheme can produce comparible results, but a problem-specific clipping is required for stable training (see Algorithm \ref{algo1} and \ref{algo2} in the Methods section). 

With this approach, network training is explicitly streamlined with the F-GMRES process. In addition, a large ensemble of residuals are created from a single structure-field pair from the training set, amplifying the effective size of the training set without the need for additional simulated data. We further note that this recurrent training scheme has the capability to learn PDE hyper-parameters implicitly. For example, our trained network is capable of solving broadband subdomain boundary value problems without explicit wavelength or resolution inputs, and the scaled information is implicitly learned from the encoding of residuals throughout the iterations. This provides a simple yet elegant solution to enable broadband model training without explicit feature or architecture engineering. Finally, the same F-GMRES loop can be utilized in both training and inference, naturally positioning the network as a general preconditioner. More details pertaining to the training process are provided in the Supplementary Information.

\begin{figure}[ht!]
\centering
\includegraphics[width=1.0\textwidth]{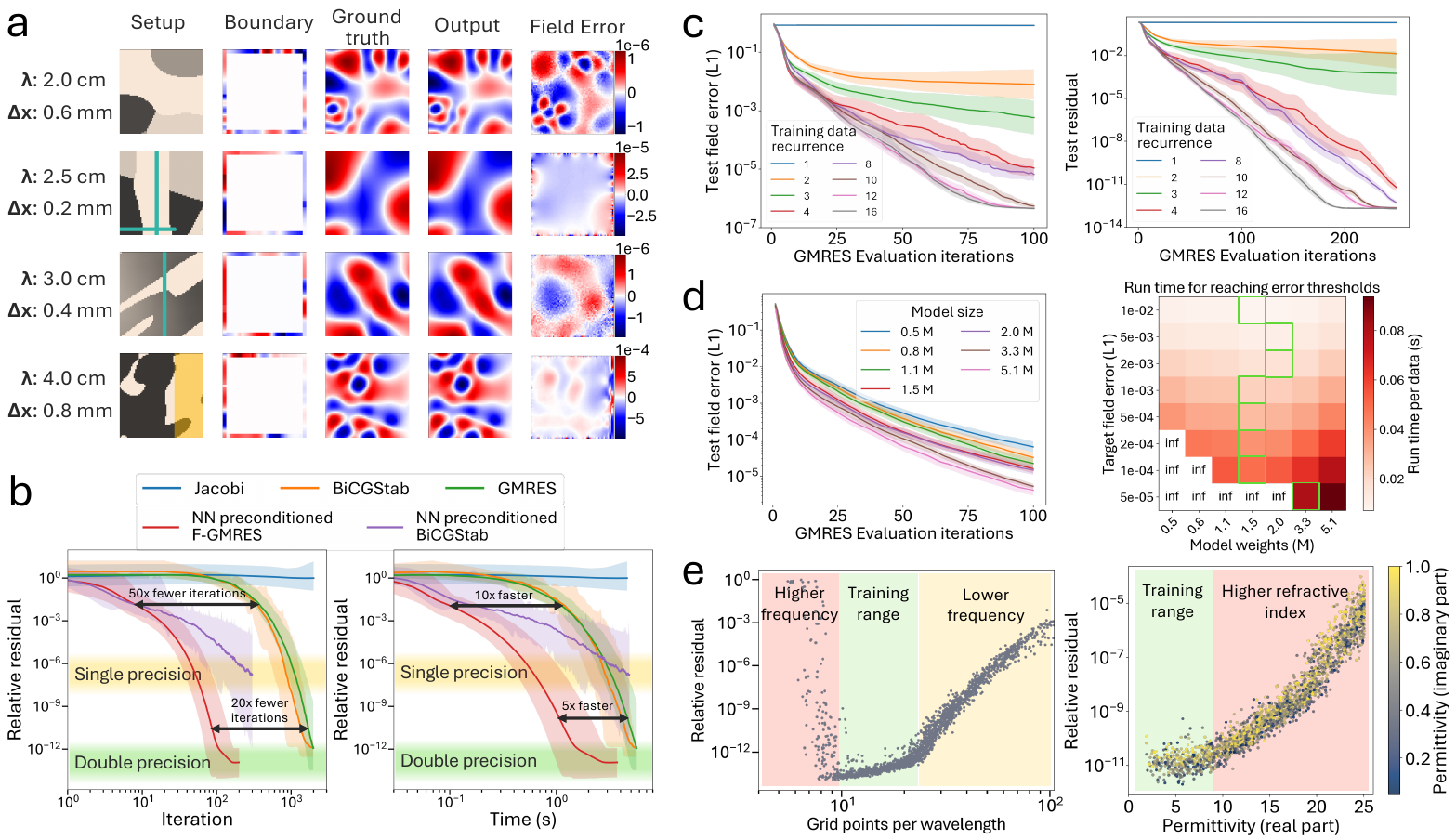}
\caption{PDE subdomain solving with network-enhanced preconditioned F-GMRES. \textbf{a} Examples of the subdomain boundary value problems showcasing heterogeneous material maps, Robin-type boundary conditions, sources (cyan), use of PML (yellow), wavelengths, and spatial resolutions. The outputs are produced with NN-preconditioned F-GMRES. \textbf{b.} Benchmarks of network-preconditioned iterative methods with conventional iterative methods. The shaded width for each line represents one standard deviation of variation as computed using 30 test data. \textbf{c.} Comparison between models with the same number of weights but trained under different numbers of training data  recurrence with GMRES. The shaded region represents one standard deviation of variation computed using 500 test data. \textbf{d.} Comparison between models with the same training setup and different number of weights. The green boxes indicate the best model size for achieving each field error threshold within the given time budget. \textbf{e.} Benchmark on out-of-distribution samples. }\label{fig3}
\end{figure}

To evaluate the efficacy of the trained subdomain models, we utilize them as preconditioners for F-GMRES and solve problems based on test data comprising arbitrary dielectric material and source distributions. Figure \ref{fig3}a shows examples of subdomain boundary value problems with heterogeneous parameter maps and random sources, PML distributions, wavelengths, and resolutions that are accurately solved using F-GMRES preconditioned by the trained model. A comparison of our approach with efficient implementations of CPU-based GMRES and BiCGSTAB using the PETSc library\cite{dalcin2011parallel} indicates that our implementation requires 20 to 50 times fewer iterations to reach a given residual threshold (Figure \ref{fig3}b), showcasing the advantage of neuro-preconditioning over conventional methods. These implementations consider a batch size of 1 for the time benchmark. Interestingly, we observe that even if the ground truth data and model weights are in single precision with 32-bit floating point numbers, the model is able to accurately solve problems with double precision when used in a double precision F-GMRES algorithm with 64-bit floating point numbers. We further find that our network can also be used as an effective preconditioner for BiCGSTAB, but due to the sensitivity of BiCGSTAB to precision, it is reliable only up to single precision. 

We further consider models that are trained with different training data recurrence (1 to 16) within the training F-GMRES loop, to quantify the impact of our iterative training scheme on preconditioner performance. The residual and field error as a function of model preconditioner iteration are shown in Figure \ref{fig3}c. We observe that as the training data recurrence in F-GMRES increases, the preconditioner also improves with faster convergence rate. Note that the training epochs for models with different training data recurrence are adjusted such that the total number of updates and training time are kept the same. This is an indication that the recurrent training scheme is effective in creating more diverse residual-error pairs to better train the model from a fixed-size dataset. 




Our neural operator preconditioner can exhibit improved performance as the network gets larger, at the expense of inference time.  This naturally raises the question of what the best model size is for a given problem: small models may take too many iterations to reach a certain error threshold while large models take too much time for each iteration.  To understand and optimize this tradeoff, we parametrically sweep the number of model weights by varying the number of Fourier modes in the linear transform and number of hidden channels, and we quantify the performance and run time of each model.  The results are summarized in Figure \ref{fig3}d and confirm the tradeoff between model size, performance, and run time.  In our experiments, models with approximately 1.5M weights are optimal. We leave further improvements in model performance via the meta-optimization of model architecture, size, and training trajectories to future research. 


An examination of how well our preconditioner performs with out-of-distribution data is summarized in Figure \ref{fig3}e, which shows F-GMRES simulation results of 2000 samples with different scales (i.e., grid points per wavelength) and dielectric distributions, including those featuring complex dielectric constants. For in-distribution cases that follow the parameter distribution of the training set, the neural network-preconditioned F-GMRES is effective in minimizing the residual with double precision. We find that the model can remain as an effective preconditioner for distributions outside of the training range, and notably, it can generalize to higher and complex-valued permittivity not seen during training. For subdomain systems featuring larger and complex-valued dielectric constants beyond the training set distribution, our solver still converges as long as the spatial frequencies within the subdomain are within the range of the training set, indicating that the network has generalized and has effectively learned Maxwell's equations. However, when the model attempts to minimize residuals with higher or lower spatial frequencies outside the range of the training set, it becomes less effective, as the learned weights in the convolution and Fourier layers are targeted for a given range of spatial frequencies. While further enlarging the training set or network capacity can help address these limitations, global-domain partitioning, discussed in the following section, presents a more scalable approach.

\subsection{Global domain solving with domain decomposition}

\begin{figure}[h]
\centering
\includegraphics[width=1.0\textwidth]{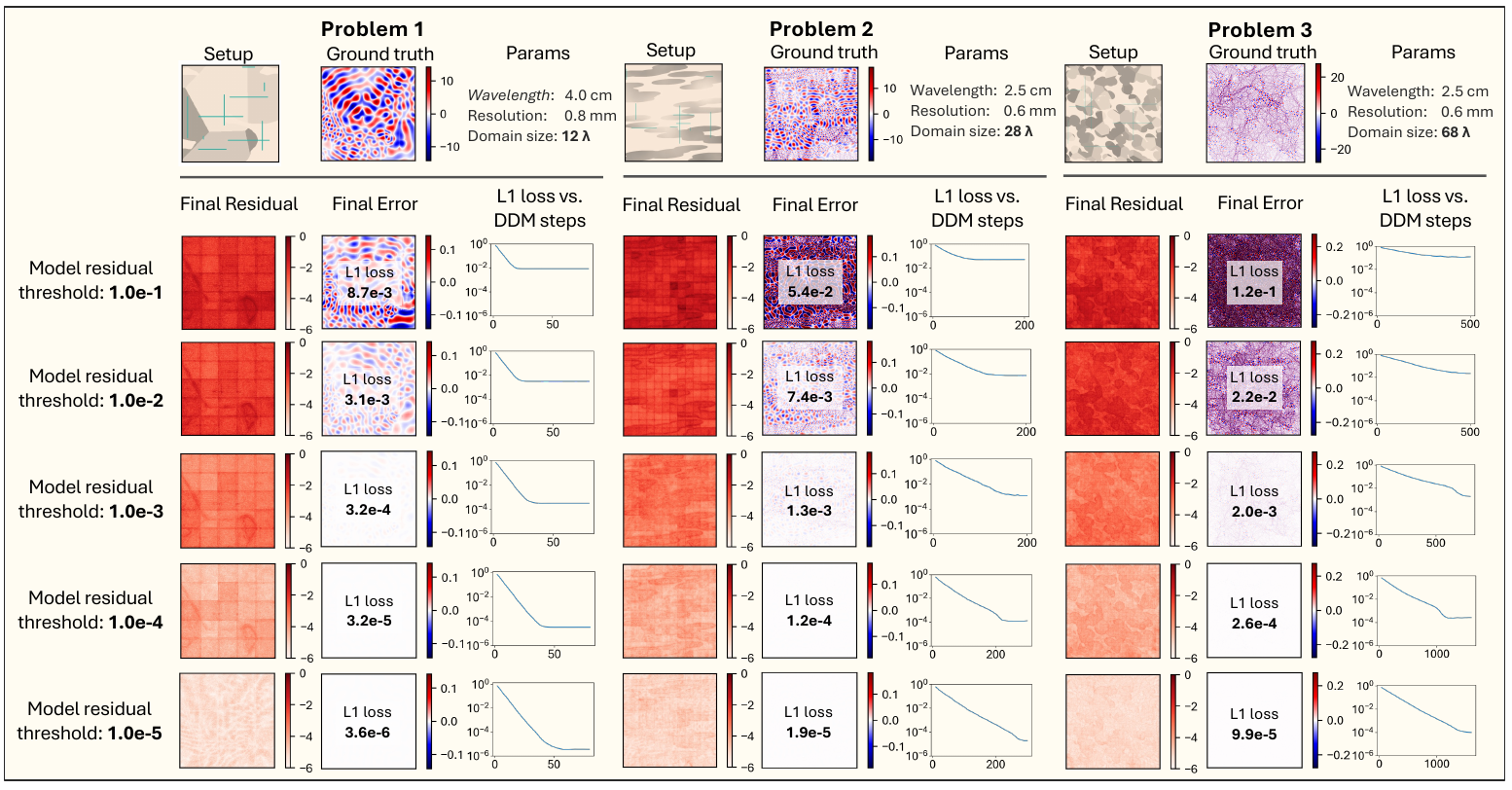}
\caption{Maxwell PDE solving with the one-level Overlapping Schwarz Method. Three problems with different wavelengths, resolutions and sizes are simulated. The problems comprise grids of $5 \times 5$, $12 \times 12$, and $25 \times 25$ subdomains. The global solutions converge with increasing accuracy as we increase the accuracy of the subdomain model by imposing stricter residual thresholds. The final field error plots are shown in a scale spanning -1\% to 1\% the maximum ground truth field value. The simulations are conducted with single precision.}\label{fig4}
\end{figure}

We now show that our subdomain preconditioner models can be utilized in domain decomposition algorithms to solve large scale electromagnetics problems. We first show that our subdomain preconditioner can combine with one-level overlapping Schwarz methods to solve global problems with machine precision.  We further quantify how subdomain model accuracy affects global solution accuracy as the problem size increases. To demonstrate, we solve a range of problems featuring random dielectric distributions, source distributions, boundary conditions, wavelengths, and sizes, all using the same subdomain preconditioner. To solve these problems, we use the approach outlined in Figure \ref{fig1}: the global problem is solved using an outer loop of domain decomposition iterations to update the subdomain boundary conditions until global convergence, and for each domain decomposition iteration, an inner loop of network-preconditioned F-GMRES subdomain iterations are performed in parallel. The inner loop stops after the F-GMRES reaches a certain predefined residual threshold. The results are summarized in Figure \ref{fig4} for different subdomain residual thresholds.  

While all setups are able to achieve convergence, the final global solution converges closer to the ground truth solution when we impose a tighter subdomain residual threshold, at the expense of needing more iterations. In addition, as the global problem size increases, more accurate subdomain solutions are required in order to obtain a globally accurate solution. In this regard, we envision that fully-convergent subdomain solvers like F-GMRES are required for true scalability to large global problems. We also want to point out that even though we stop the subdomain F-GMRES iterations when it reaches a certain residual threshold in this benchmark, in practice we found out that the global solution can also converge even if the subdomain is not converged at each iteration. For example, even if we only perform 1 GMRES iteration for the inner loop, the global outer loop can still converge, often with less total time. However, the stability is worsened and breakdowns in convergance can occur. We leave the meta-optimization of subdomain and global residual minimization, as well as their robustness, to future work.

\begin{figure}[ht!]
\centering
\includegraphics[width=1.0\textwidth]{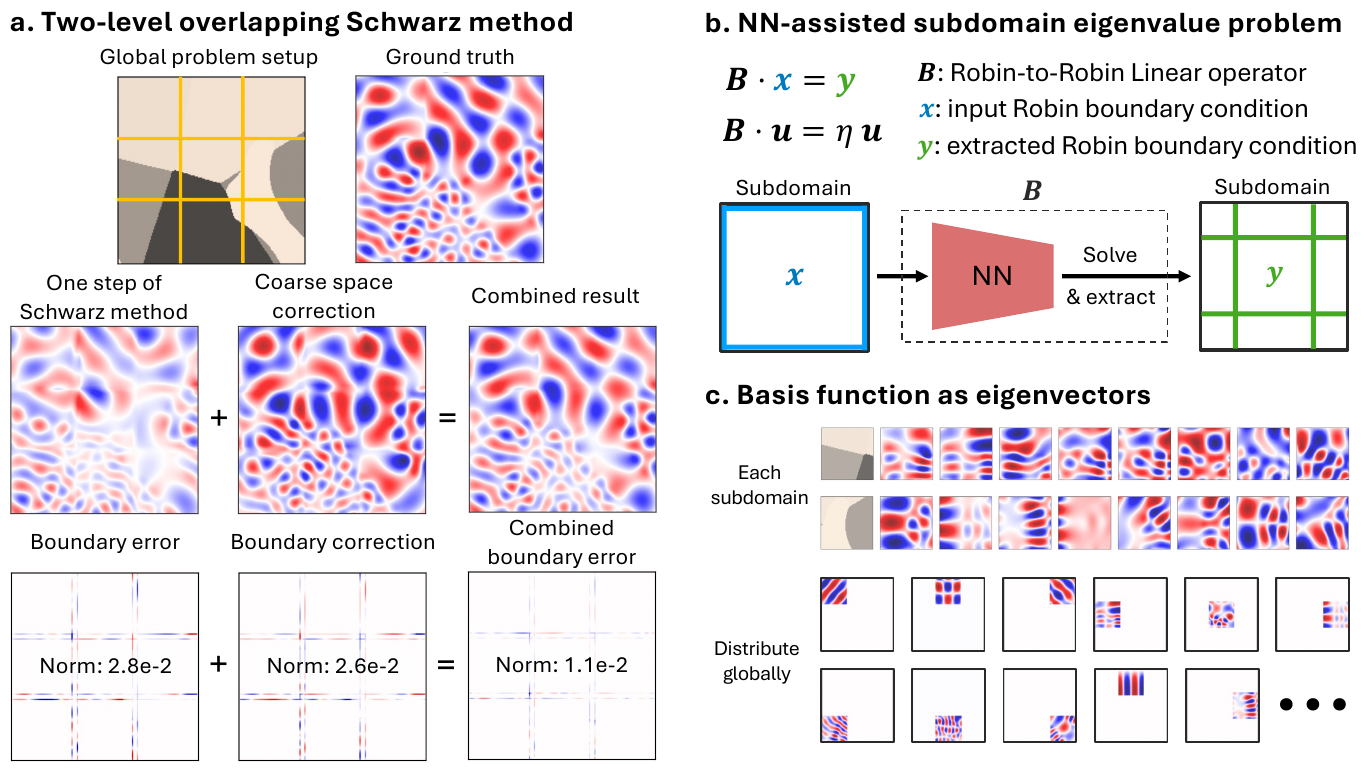}
\caption{Two-level overlapping Schwarz Method and coarse space. \textbf{(a)} Exemplary problem with 3 by 3 subdomains. Within one iteration, the global coarse space correction amends the boundary mismatches resulting from the local solve, resulting in an accurate combined solution. \textbf{(b)} The linear operator and eigenvalue problem in the subdomain. \textbf{(c)} Examples of basis functions as eigenvectors, for each subdomain with heterogeneous parameter map. The basis functions are then distributed globally as a basis function in the coarse space.}\label{fig5}
\end{figure}

While the one-level Schwarz method converges when using accurate subdomain solutions, it ultimately requires very large numbers of domain decomposition iterations for large global domain sizes. This scaling trend is intrinsic to the one-level Schwarz method where subdomain boundary conditions propagate and update only for nearest neighboring subdomains in an iteration, thereby requiring increasing numbers of iterations for information to propagate across the global domain as the domain size increases. To overcome these limits, a means of global information exchange is required.  One typical approach involves the construction of a global coarse space, which is a low-dimensional subspace spanning the global domain that is obtained by solving a dimensionality-reduced global problem\cite{dolean2015introduction}. Coarse space solutions can be incorporated with the one-level Schwarz method to yield the two-level overlapping Schwarz method framework. In a given iteration, the coarse space correction is efficiently computed to globally correct the subdomain boundary mismatch errors from the one-level Schwarz solution (Figure \ref{fig5}a), accelerating domain decomposition convergence. 

Traditional spectral coarse space construction relies on solving eigenvalue problems for each subdomain followed by solving a dimensionality-reduced global problem \cite{nataf2010two, haferssas2015robust}, and large computational overhead is required to solve these eigenvalue problems from scratch  \cite{dolean2015introduction}. We propose to use our subdomain models to accelerate the solving of these subdomain-scale eigenvalue problems. The concept is illustrated in Figure \ref{fig5}b. The eigenvalue problems for each subdomain are defined by the linear operator $\mathbf{B}$, which acts on a given Robin-type boundary value vector $x$ and produces the boundary values for its neighbors $y$. Eigenvalue solutions to $\mathbf{B}$ comprise eigenvectors $u$ and associated eigenvalues $\eta$, and they represent electromagnetic modal solutions with boundaries $u$ and fields that can be readily evaluated using our neural operator. Modes with the largest eigenvalues $\eta$ decay the slowest, and for the purposes of coarse space construction, we consider coarse space solutions to comprise linear combinations of a finite number of slowly converging modes with a lower eigenvalue threshold. To calculate $u$ and $\eta$ for a given subdomain, we evaluate $(x,y)$ pairs using our subdomain neural operator, where $x$ are heuristically chosen as pseudo-plane waves, and we use the Rayleigh–Ritz method to obtain the approximated $u$ and $\eta$, with details in the Methods section. Calculation of the modes only needs to be performed once for a given global domain layout. This process enables automatic coarse space basis function construction, even for subdomains with highly heterogeneous dielectric distributions. The method is also accelerated with our subdomain neural operator because our subdomain models can perform fast batched evaluation of $(x,y)$ pairs for multiple subdomains in parallel.

Our process for performing coarse space correction for a given domain decomposition iteration is standard: boundary error from the one-level Schwarz method is evaluated and  linear combinations of modes that can minimize this error are determined (Figure \ref{fig5}C). For $n$ subdomains and $k_{coarse}$ modes per subdomain, this coarse problem reduces to solving a $n \times k_{coarse}$ dimensional linear problem, which in most cases are very small and can be efficiently and quickly solved using direct methods. The fields from the one-level Schwarz method and coarse space correction are combined and used for the subsequent domain decomposition iteration. To study the scaling trend of this coarse space problem, we evaluate the two-level overlapping Schwarz Method on two subdomain model choices: one with 64 by 64 subdomain size and the other with 256 by 256 subdomain size. We perform both one-level and two-level overlapping Schwarz Methods on problems with different global domain sizes, and the results are summarized in Table \ref{tab1} and Table \ref{tab2}. 

Both models are able to solve arbitrary-sized problems with different wavelengths and resolutions up to grid size of about 3000 by 3000, and physical size of up to 200 wavelengths, while the two-level Schwarz method enables a significant reduction in domain decomposition iterations. In fact, when the coarse space is well-constructed, the number of iterations to converge either stays constant or grows very slowly as the global problem size increase\cite{farhat2000two}. The term $k_{coarse}$ indicates the number of basis functions that we evaluate in each subdomain, and it is effectively equivalent to running an additional $k_{coarse}$ global domain decomposition iterations compared to the one-level Schwarz method. This one-time cost for a specific dielectric distribution is especially efficient in cases where multiple source configurations are evaluated on the same dielectric distribution. We emphasize that the effectiveness of the coarse space is crucial and remains an active research area for systems including Maxwell's equations with heterogeneous medium. In our experiments, highly resonant subdomains with discrete material boundaries worsen the effectiveness of the coarse space, and with larger subdomain size and higher frequency (fewer grids per $\lambda$), the required size of the coarse space also grows accordingly. When the heterogeneous dielectric map is grayscale and varies smoothly, the coarse space appear to be very effective, as shown in Table \ref{tab1} and Table \ref{tab2}. We have included one visual example of the convergence history in the Supplementary Information.

\begin{table}[h] 
\caption{Comparison between one-level and two-level Schwarz methods, for \textbf{64 by 64} grid subdomain preconditioner model.}\label{tab1}%
\setlength{\tabcolsep}{3pt}
\begin{tabular}{@{}cc||c|c||c|c||c|c@{}}
\toprule
\multicolumn{2}{c||}{\textbf{subdomain size}} & \multicolumn{2}{c||}{grids per $\lambda = \mathbf{80}$} & \multicolumn{2}{c||}{grids per $\lambda = \mathbf{44}$} & \multicolumn{2}{c}{grids per $\lambda = \mathbf{24}$} \\
\multicolumn{2}{c||}{\textbf{64 by 64}} & \multicolumn{2}{c||}{$k_{coarse} = \mathbf{16}$} & \multicolumn{2}{c||}{$k_{coarse} = \mathbf{24}$} & \multicolumn{2}{c}{$k_{coarse} = \mathbf{36}$} \\
\cmidrule(lr){3-4}\cmidrule(lr){5-6}\cmidrule(lr){7-8}
\makecell{\# of\\subdomains} & global size & \makecell{physical size\\ $k_{max}$ / $N_{\lambda}$} & \makecell{iterations\\1L / 2L} & \makecell{physical size\\ $k_{max}$ / $N_{\lambda}$} & \makecell{iterations\\1L / 2L} & \makecell{physical size\\ $k_{max}$ / $N_{\lambda}$} & \makecell{iterations\\1L / 2L} \\
\midrule
 4 by 4    & (240, 240)   & 25 / 3.1   & \textbf{ 22 / 7 }   & 46 / 5.7 &  \textbf{ 26 / 5 } & \hspace{1.5pt} 87 / 10.7 & \textbf{\hspace{1.5pt} 24 / 4}\\
 8 by 8    & (480, 480)   & 50 / 5.8   & \textbf{ 48 / 9 }  & \hspace{1.5pt} 92 / 10.7 & \textbf{ 69 / 7 }  & 172 / 20.1 & \textbf{\hspace{1.5pt} 67 / 5}  \\
 12 by 12  & (720, 720)   & 75 / 8.8   & \textbf{\hspace{1.5pt} 75 / 10}  & 137 / 16.2 & \textbf{107 / 8 \hspace{1.5pt}} & 256 / 30.5 & \textbf{117 / 6} \\
 16 by 16  & (960, 960)   & 100 / 12.3 & \textbf{102 / 10}  & 183 / 22.7 & \textbf{163 / 8 \hspace{1.5pt}} & 343 / 42.6 & \textbf{155 / 6}\\
 24 by 24  & (1440, 1440) & 150 / 18.4 & \textbf{181 / 19} & 274 / 33.9 & \textbf{284 / 10} & 513 / 63.6 & \textbf{271 / 7}\\
\botrule
\end{tabular}
\begin{tablenotes}
\item $k_{max}$: maximum wavenumber, with domain size normalized to 1. 
\item $N_{\lambda}$: simulation size in terms of number of wavelengths within the global domain.
\item 1L / 2L: short for one-level and two-level
\end{tablenotes}
\vspace{0.3cm}
\caption{Comparison between one-level and two-level Schwawrz methods, using a \textbf{256 by 256} grid subdomain preconditioner model.}\label{tab2}%
\setlength{\tabcolsep}{3pt}
\begin{tabular}{@{}cc||c|c||c|c||c|c@{}}
\toprule
\multicolumn{2}{c||}{\textbf{subdomain size}} & \multicolumn{2}{c||}{grids per $\lambda = \mathbf{28}$} & \multicolumn{2}{c||}{grids per $\lambda = \mathbf{22}$} & \multicolumn{2}{c}{grids per $\lambda = \mathbf{15}$} \\
\multicolumn{2}{c||}{\textbf{256 by 256}} & \multicolumn{2}{c||}{$k_{coarse} = \mathbf{100}$} & \multicolumn{2}{c||}{$k_{coarse} = \mathbf{120}$} & \multicolumn{2}{c}{$k_{coarse} = \mathbf{160}$} \\
\cmidrule(lr){3-4}\cmidrule(lr){5-6}\cmidrule(lr){7-8}
\makecell{\# of\\subdomains} & global size & \makecell{physical size\\ $k_{max}$ / $N_{\lambda}$} & \makecell{iterations\\1L / 2L} & \makecell{physical size\\ $k_{max}$ / $N_{\lambda}$} & \makecell{iterations\\1L / 2L} & \makecell{physical size\\ $k_{max}$ / $N_{\lambda}$} & \makecell{iterations\\1L / 2L} \\
\midrule
 4 by 4    & (994, 994)   & 299 / 35.4   & \textbf{\hspace{2pt} 44 / 4 \hspace{1.5pt}}   & \hspace{1.5pt} 392 / 46.5 &  \textbf{ 40 / 5 } & \hspace{1.5pt} 573 / 67.9 & \textbf{50 / 5}\\
 6 by 6    & (1486, 1486)   & 453 / 53.8   & \textbf{\hspace{2pt} 91 / 4 \hspace{1.5pt}}  & \hspace{1.5pt} 594 / 70.7 & \textbf{ 75 / 6 }  & 868 / 103 & \textbf{79 / 9}  \\
 8 by 8  & (1978, 1978)   & 592 / 70.9   & \textbf{170 / 7 \hspace{1.5pt}}  & \hspace{1.5pt} 777 / 93.0 & \textbf{174 / 9 \hspace{1.5pt}} & 1130 / 136 & \textbf{138 / 12} \\
 10 by 10  & (2470, 2470)   & 752 / 89.5 & \textbf{262 / 8 \hspace{1.5pt}}  & \hspace{1.5pt} 988 / 117\hspace{1.5pt} & \textbf{224 / 11} & 1440 / 172 & \textbf{174 / 16}\\
 12 by 12  & (2962, 2962) & 887 / 106\hspace{2pt} & \textbf{421 / 15} & 1160 / 140\hspace{1pt} & \textbf{350 / 19} & 1700 / 203 & \textbf{251 / 23}\\
\botrule
\end{tabular}
\end{table}


Finally, we demonstrate that our general purpose multi-level solver can be used for accurate and scalable inverse design using gradient-based optimization. Freeform density-based optimization is performed using the adjoint variables method (AVM) coupled with gradient-based optimization. We begin with a homogeneous grayscale dielectric representing our photonic device, and we use Adam-based gradient descent to iteratively push the dielectric values at each voxel in a manner that maximizes our Figure of Merit. Differentiable filtering and projection operations can be used to push the dielectric distribution towards increasingly binary values over the course of optimization. Gradients to the dielectric constant at each voxel are calculated using forward and adjoint simulations with our solver, and in cases where the gradients from AVM are part of a larger computational graph, a custom Jacobian vector product (JVP) ensures the full computational graph is differentiable. We consider implicit differentiation over automatic differentiation because the computational graph quickly becomes intractable for large problems, rendering automatic differentiation infeasible due to memory constraint issues.


\begin{figure}[ht!]
\centering
\includegraphics[width=1.0\textwidth]{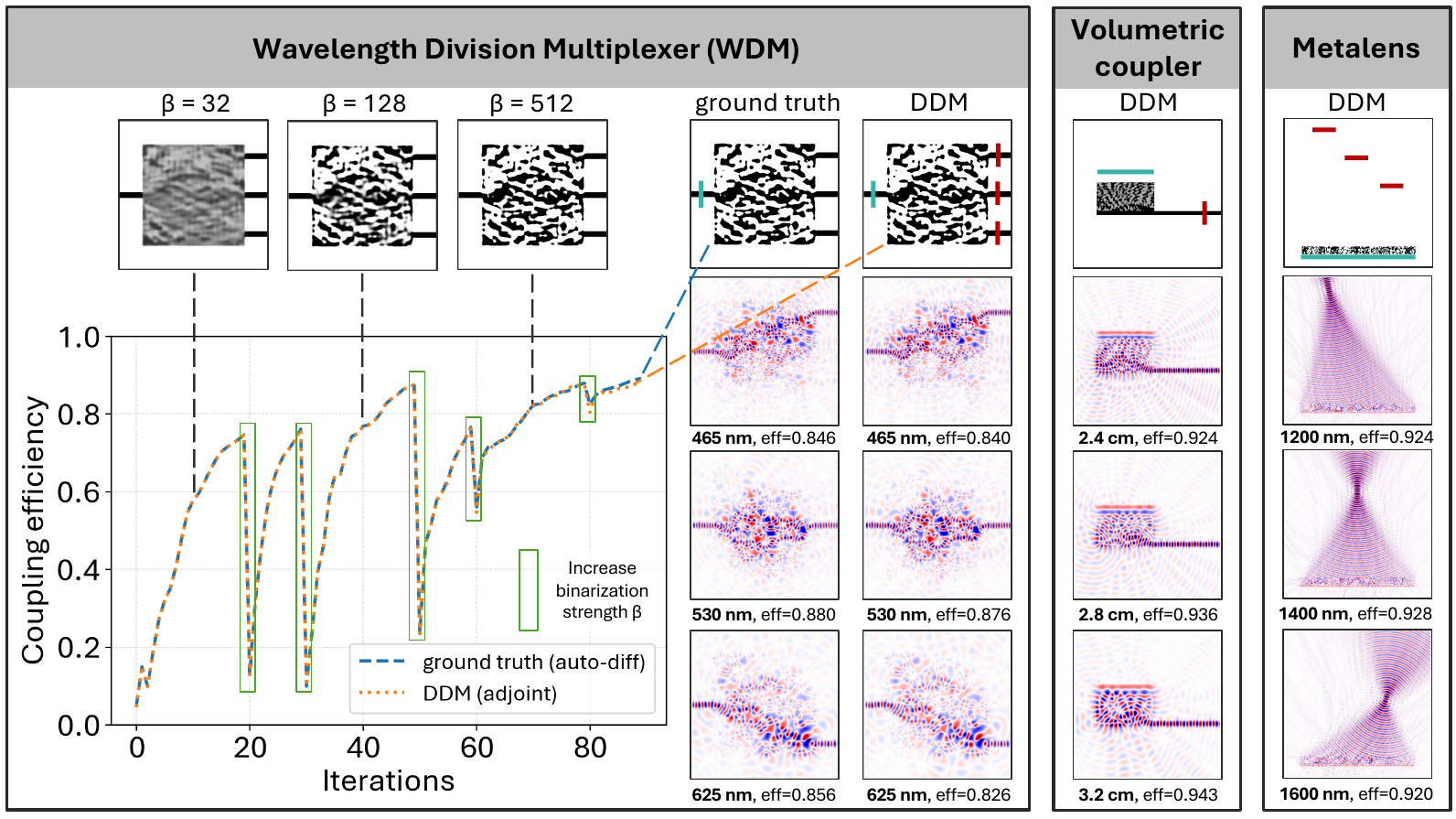}
\caption{Freeform gradient-based inverse design with our multi-level solver.  Topology optimization is performed on three types of photonic devices using the adjoint variables method and implicit differentiation: a visible light wavelength division multiplexer (left), a multi-wavelength radio frequency coupler (center), and a multi-functional volumetric near-infrared metalens (right). The cyan region indicates the forward source region and the red regions indicate the figure-of-merit windows, which is also the location for adjoint sources. The optimization trajectories and final freeform devices based on our neural network solver are nearly identical to the ground truth. A single subdomain model of input size $256 \times256$ is used in all three examples.}\label{fig6}
\end{figure}

Results for the freeform design of a wavelength division multiplexer, multi-wavelength volumetric coupler, and multi-functional metalens are shown in Figure \ref{fig6}.  The multiplexer and metalens are designed to have binary dielectric values while the coupler is designed to have grayscale dielectric values. The subdomain solutions are specified to have high accuracy (i.e., residual $< 10^{-6}$). In all cases, the devices exhibit high efficiencies typical of freeform volumetric photonic media. A comparison of the multiplexer optimization trajectory produced using our solver and a ground truth FDFD solver, which uses automatic differentiation to calculate gradients, are nearly identical.  The final devices optimized from both methods have nearly identical layouts and performance, indicating that our solver performs nearly identically to a ground truth FDFD solver. To the best of our knowledge, these results represent the first demonstrations of photonic inverse design for large scale domains ($>100 \lambda$) with a neural network surrogate solver.

\section{Discussion}
In this work, we demonstrate for the first time a scalable and accurate approach to building general purpose PDE surrogate solvers using neural networks. The key to our approach is to build a subdomain surrogate solver that is capable of solving a general boundary value problem for a diverse range of wavelengths, source distributions, and auxiliary parameters. Utilization of this subdomain solver as a subdomain preconditioner for F-GMRES and as an eigenvalue mode solver for course grid construction enables streamlining with multilevel domain decomposition methods and the solving of subdomain and global domain problems with machine precision.

While our approach points out a promising path forward in neural PDE solving, there are many interesting directions for future research and opportunities to improve our concept. First, there are many opportunities to explore and understand how the model architecture, model size, and its implementation in domain decomposition workflows can be optimized to accelerate PDE solving. New architecture concepts beyond our modified FNO have the potential to utilize fewer network weights without a reduction in performance, pushing subdomain capabilities beyond the networks optimized in Figure \ref{fig3}d. The optimization of subdomain size in the context of multilevel PDE solving is important to navigate the tradeoff between subdomain size, model complexity, coarse space construction speed, and domain decomposition convergence. There also are opportunities to modulate the accuracy of subdomain and global solutions over the course of global iterative solving in a manner that reduces computational cost. For example, an intuitive strategy involves the gradual increase of subdomain accuracy over the course of global iterative solving such that the subdomain solutions in early iterations are approximate but sufficiently accurate to facilitate global boundary updating. 

Second, a detailed computational cost and speed analysis is required to understand where our neural network solvers have an opportunity to outperform conventional solvers. The problems that we consider in this study are the two-dimensional problems with up to several million voxels, for which efficient direct solvers based on sparse matrix inversion can apply. In this limit, when benchmarked against a highly efficient direct sparse matrix solver (PARDISO\cite{schenk2001pardiso}) running on a workstation with 80 CPU cores, our solver (running on 1 NVIDIA RTX A6000 GPU) is approximately twice as slow. We anticipate that for problems with larger sizes where direct solvers cannot apply (i.e., $>$ 10 million grids in 2D and $>$ 1 million voxels in 3D), our methods will have a decisive advantage, and proper benchmarking with fully optimized workflows will be required to understand these computational scaling trends.



Third, research is required to understand how to stably scale these methods to domain sizes much larger than $200\lambda$, particularly for hyperbolic PDE problems. Domain decomposition methods were originally developed for elliptic PDEs, which are positive semi-definite, and it remains an open question how these methods can stably converge and scale when applied to indefinite systems and hyperbolic PDEs such as Maxwell's equations. While we were able to obtain convergent results in this study, we did also observe that as the problem size grew, convergence became more and more sensitive to hyperparameters, such as the amount of overlap between subdomains and the momentum term in the boundary update step. The effectiveness of the coarse space also slowly degraded as the problem size increased. We anticipate that neural networks specifically trained to perform  coarse space computation can potentially help with domain size scaling.  We also anticipate that other types of domain decomposition methods, including Finite Element Tearing and Interconnecting and Multigrid methods, could potential offer certain benefits in terms of efficiency and stability.  

\section{Methods}\label{sec:Methods}
\subsection{Training data generation}
Both physics-only and data-driving training data generation schemes are used in this work. For the model with 64 by 64 subdomain problem size, ground truth simulations are conducted with an FDFD solver\cite{hughes2019forward}. We generate diverse dielectric distributions using Gaussian random field and Voronoi map masked by free-from shapes, to create continuous and discrete material distributions encountered in nanophotonics inverse designs. The Gaussian random fields and Voronoi maps have feature sizes from 1 to 50 pixels to cover structures from subwavelength to several-wavelength large. We run 4,000 simulations with grid size 600 by 600 on these dielectric data, with randomly placed line sources with diverse sinusoidal profiles, under wavelength from 2 cm to 4 cm, resolution from 0.2 mm to 0.8 mm, and permittivity from 1 to 8. The generated field maps, dielectric distributions, source and PML maps are cropped into a total of 1M training samples with size of 64 by 64. The Robin-type boundary values are prepared using the cropped dielectric maps and the fields, which are used to compute the residual maps used during training.

For the model with 256 by 256 subdomain problem size, random data is generated for physics-only training, without running ground truth simulations. The same methods of Gaussian random field and Voronoi map are used to generate dielectric distributions. The source maps are used as pure gaussian noise that span the whole domain, which is crucial for the convergence of the physics-only training. The PMLs are randomly placed in both x and y directions to cover scenarios encountered in DDM subdomains. The Robin-type boundary values are also generated as Gaussian noise with different length scales (standard deviation). The source and boundary values are heuristically scaled down within the PML regions for more stable training.

In both cases, a uniaxial-PML (UPML) is adopted. Which has the formulation:
\begin{align}
\nabla \times \mathbf{E} = -i\omega \mu [s] \mathbf{H} \\
\nabla \times \mathbf{H} = \mathbf{J} + i\omega \varepsilon [s] \mathbf{E}
\end{align}
in which the tensor $[s]$ has form:
\begin{align}
[s] = \begin{bmatrix}
\cfrac{s_ys_z}{s_x}, &0, &0 \\
0, &\cfrac{s_zs_x}{s_y}, &0 \\
0, &0, &\cfrac{s_xs_y}{s_z} 
\end{bmatrix}
\end{align}
with 
\begin{align}
s_\zeta(\zeta) = 1 + \cfrac{\sigma_\zeta(\zeta)}{i\omega \varepsilon_0}, \quad \sigma_\zeta(\zeta) = \sigma_{\zeta, max} \cdot \Big(\cfrac{\zeta}{d}\Big)^{m}, \quad \sigma_{\zeta, max} = -\cfrac{(m+1)\ln R}{2\eta_0 d\zeta}, \quad \zeta = x, y, z
\end{align}
in which $d$ is the width of the PML, R is the ideal reflectivity (chosen as $e^{-30}$), and $\eta_0$ is the impedance of free space, and m is a coefficient chosen as 3. The s matrices (in our case $s_x$ and $s_y$) are prepared and cropped as auxiliary inputs to the model.

\subsection{Subdomain training and inference algorithms}
Here we present the NN-preconditioned Flexible GMRES algorithm for training and inference. Training is performed under single precision and both single and double precision can be used for inference. Both physics-only and physics-plus-data loss functions can be utilized to train the network. The network gradients are updated in each GMRES iteration. We want to mention that back-propagation through the entire GMRES iterations is also possible, but in our practice we observe more stable training with back-propagation at each step.

For the ease of implementation, we also present a training method based on fixed point iterations, which gives near identical result compared to training with GMRES. One difference is that fix-point-iterations are divergent for indefinite systems, especially at the beginning of the training where network produces noisy outputs. We found out that a clipping strategy is effective to obtain stable training. The exact clipping threshold is dependent on the data recurrence $N$, as well as the PDE. 
\begin{algorithm}
\caption{NN-preconditioned Flexible GMRES (NN-FGMRES)}\label{algo1}
\begin{algorithmic}[1]
\Require Model $\mathbf{M}$, right hand side $\mathbf{b}$, auxiliary input $\mathbf{aux}$, operator $\mathbf{A}$, residual threshold $\mathbf{r_{th}}$ or maximum data recurrence \textbf{N}
\Require \textbf{(in training)} MSE loss function $\mathbf{L_2}$, optimizer \textbf{opt}, weight for data and residual loss $\mathbf{c_d}$, $\mathbf{c_r}$, ground truth field \textbf{y} (optional)
\State$\mathbf{x_0} = \mathbf{0}$
\State$\mathbf{r_0} = b - \mathbf{A}(\mathbf{x_0})$
\State$\beta = ||\mathbf{r_0}||_2$
\State$\mathbf{v_1} = \mathbf{r_0}/\beta$
\For{$j=1,...,N$} 
        \State {\color{blue} opt\text{.}zero\_grad()} \Comment{{\color{blue} Training only}}
        \State $\mathbf{z_j} = \mathbf{M}(\mathbf{v_{j}}, aux)$ \Comment{{\color{orange} Inference: convert input to 32 bit, output back to 64 bit}}
        \State $\mathbf{w} = \mathbf{A}(\mathbf{z_j})$
        \For{$i=1,...,j$}
                \State $h_{i,j} = (\mathbf{w}, \mathbf{v_i})$ \Comment{complex inner product}
                \State $\mathbf{w} = \mathbf{w} - h_{i,j}\mathbf{v_j}$
        \EndFor
        \State $h_{j+1,j} = ||\mathbf{w}||_2$
        \State $\mathbf{v_{j+1}} = \mathbf{w} / h_{j+1,j}$
        \State $\mathbf{Z_j} := [\mathbf{z_1},...,\mathbf{z_j}]$, $\mathbf{\overline{H}_j} = \{h_{i,k}\}_{1\le i\le j; 1\le k\le j}$
        \State $\mathbf{y_j} = argmin_\mathbf{y} ||\beta \mathbf{e_1} - \mathbf{\overline{H}_j} y||_2 $
        \State $\mathbf{x_j} = \mathbf{x_0} + \mathbf{Z_j}\mathbf{y_j}$
        \State $\mathbf{r_j} = \mathbf{b} - \mathbf{Aop}(\mathbf{x_j})$
        \State {\color{blue}$loss = c_r\mathbf{L_2}(\mathbf{r_j}) + (optional)\ c_d\mathbf{L_2}(\mathbf{x}-\mathbf{y})$}  \Comment{{\color{blue} Training only}}
        \State {\color{blue}loss\text{.}backward(); opt\text{.}step()} \Comment{{\color{blue} Training only}}
        \State {\color{blue}$stop\_gradient(\mathbf{v_j}, \mathbf{Z_j}, \mathbf{H_j})$ }\Comment{{\color{blue} Training only}}
        \If{$\mathbf{r_j} < r_{th}$}\Comment{{\color{orange} Inference only}}
            \State break\Comment{{\color{orange} Inference only}}
        \EndIf\Comment{{\color{orange} Inference only}}
\EndFor
\end{algorithmic}
\end{algorithm}

\begin{algorithm}
\caption{Fixed-point-iteration training of NN-preconditioner}\label{algo2}
\begin{algorithmic}[1]
\Require Model $\mathbf{M}$, right hand side $\mathbf{b}$, auxiliary input $\mathbf{aux}$, operator $\mathbf{A}$, maximum data recurrence \textbf{N}, clipping threshold $\mathbf{th_{clip}}$, MSE loss function $\mathbf{L_2}$, optimizer \textbf{opt}, weight for data and residual loss $\mathbf{c_d}$, $\mathbf{c_r}$, (optional) ground truth field \textbf{y}
\State$\mathbf{x} = \mathbf{0}$
\State$\mathbf{r} = \mathbf{b} - \mathbf{A}(\mathbf{x})$
\For{$j=1,...,N$} 
        \State opt\text{.}zero\_grad()
        \State $\mathbf{e} = \mathbf{M}(\mathbf{r}, aux)$ 
        \State $\mathbf{x} = \mathbf{x} + \mathbf{e}$
        \State $\mathbf{r} = \mathbf{b} - \mathbf{A}(\mathbf{x})$
        \State $loss = c_r\mathbf{L_2}(\mathbf{r}) + (optional)\ c_d\mathbf{L_2}(\mathbf{x}-\mathbf{y})$
        \State loss\text{.}backward(); opt\text{.}step()
        \State $r = stop\_gradient(clip(\mathbf{r}, th_{clip}))$
\EndFor
\end{algorithmic}
\end{algorithm}

\subsection{Two-level Schwarz method with coarse space}
All of the one-level and two-level overlapping Schwarz methods used in this work are implemented with an abstraction which we call the ``partitioner". For a given global problem configuration, the partitioner takes care of partitioning global domain data into overlapping (or non-overlapping) subdomain data, and assembling them back to the global domain. The subdomains all have a fixed grid size (so they can be efficiently processed as a batch), but they can in principle have different grid resolutions and thus different physical sizes (as shown in Figure \ref{fig1}a). For example, a natural strategy can be a quad-tree partition of the global domain, and whenever the physical size of a quadrant is larger than a chosen threshold (e.g. $5\lambda$), it will be decomposed into 4 smaller regions. The subdomains can be arbitrarily arranged and share different overlap with its neighbors, as long as their union area covers the whole global domain. In Figure \ref{fig4}, \ref{fig5} and \ref{fig6} and table \ref{tab1}, \ref{tab2}, the subdomains are arranged in a rectangular grid, with each row and column have the same overlapping configurations. Once a specific partition geometry is chosen, at each domain decomposition iteration, the partitioner will take care of the boundary update for each subdomain from its neighbors, as well as a coarse space correction in two-level methods. The subdomain data will be batched and iteratively solved with the NN-preconditioned F-GMRES algorithm.

For the coarse space construction, it is well known that for subdomains with homogeneous medium, the natural basis are plane waves traveling in different directions spanning the unit circle\cite{farhat2000two}. We use the Rayleigh–Ritz method and heuristically choose a set of pseudo-plane wave trial basis as probe boundary conditions to construct a subspace that approximates the Robin-to-Robin map. The pseudo plane wave boundary conditions are constructed by first compute the wave vector inside the medium for a chosen plane wave direction, and then accumulate the phase around the boundary. We have also experimented with other trial boundary conditions such as random Gaussian or Fourier modes at the boundary, which produces acceptable but worse results. Note that when solving for each boundary value problem for these RtR maps, homogeneous equations without the source term are used. We have also tried the singular value decomposition method to get the basis vectors, and it yields similar quality coarse space compared to solving for the eigenvectors. At each domain decomposition iteration, we solve the coarse space correction as a linear system:
\begin{align}
BQz_{coarse} = e_{bc}
\end{align}
in which $B$ is the Robin-to-Robin map, each column of $Q$ is the boundary values for a basis vector, $z_{coarse}$ is the linear coefficients of the basis we want to solve, which minimizes the boundary mismatch error $e_{bc}$. The standard practice is to solve for its normal equation:
\begin{align}\label{coarse_eq}
(\overline{BQ})^TBQz_{coarse} = (\overline{BQ})^T e_{bc}
\end{align}
for which we can assemble (and factor if needed) the small $(\overline{BQ})^TBQ$ matrix beforehand for fast solve at run time. The process is summarized in Algorithm \ref{algo3}.

\begin{algorithm}
\caption{Two-level Schwarz method with coarse space}\label{algo3}
\begin{algorithmic}[1]
\Require Dielectric distribution $\varepsilon$, source distribution $\mathbf{j}$, PML distribution $\mathbf{S}$, subdomain partition configuration \textbf{config}, number of basis functions per subdomain $\mathbf{k_{coarse}}$, subdomain solver $\mathbf{M}$, subdomain residual threshold $\mathbf{th_{sub}}$, global residual threshold $\mathbf{th_{global}}$, momentum $m$
\vspace{1em}
\State $\mathbf{P} = init\_partitioner(\varepsilon, \text{config})$
\State eps\_batch, src\_batch, PML\_batch = P.partition($\varepsilon$, $\mathbf{j}$, $\mathbf{S}$)
\State \textbf{M}.setup(eps\_batch, PML\_batch) \Comment{zero source}
\vspace{1em}
\If{use\_coarse\_space} \Comment{One-time pre-compute for coarse space}
    \State $\mathbf{x} = prepare\_pseudo\_plane\_wave\_bc(k_{coarse})$
    \State $\mathbf{x} = gram\_schmidt(\mathbf{x})$
    \State $\mathbf{fields} = \mathbf{M}.solve(\mathbf{x}, th_{sub})$
    \State $\mathbf{y} = \mathbf{P}.extract\_neighboring\_bc(\mathbf{fields})$
    \State $H = (\mathbf{x}, \mathbf{y})$ \Comment{complex inner product}
    \State $\eta, \mathbf{u} = eig(H)$
    \State $\mathbf{P}.assemble\_BQTBQ(\mathbf{u})$
\EndIf
\vspace{1em}
\State $\mathbf{x}, \mathbf{r} = 0, 0$
\State $bc = \mathbf{P}.get\_bc(\mathbf{x\_batch})$
\State \textbf{M}.setup(eps\_batch, src\_batch, PML\_batch) 
\While {$r>th_{global}$} \Comment{Domain decomposition loop}
    \State $\mathbf{x}_{new} = \mathbf{M}.solve(bc, th_{sub})$
    \State $bc_{new} = \mathbf{P}.get\_bc(\mathbf{x}_{new})$
    \State $bc = m\cdot bc + (1-m)\cdot bc_{new}$
    \State $r = get\_residual(\mathbf{P}.assemble(\mathbf{x}_{new}))$
    \If{use\_coarse\_space}
        \State $e_{bc} = \mathbf{P}.get\_bc\_mismatch(\mathbf{x}_{new})$
        \State $z_{coarse} = \mathbf{P}.solve\_coarse\_space(e_{bc})$ \Comment{Solve for equation \ref{coarse_eq}}
        \State $bc_{coarse} = \mathbf{P}.coarse\_space\_correction(z_{coarse})$
        \State $bc = bc + bc_{coarse}$
    \EndIf
\EndWhile
\vspace{1em}
\State \textbf{Return} $\mathbf{P}.assemble(\mathbf{x}_{new})$
\end{algorithmic}
\end{algorithm}




\newpage
\bibliography{sn-article}


\begin{thebibliography}{41}
\ifx \bisbn   \undefined \def \bisbn  #1{ISBN #1}\fi
\ifx \binits  \undefined \def \binits#1{#1}\fi
\ifx \bauthor  \undefined \def \bauthor#1{#1}\fi
\ifx \batitle  \undefined \def \batitle#1{#1}\fi
\ifx \bjtitle  \undefined \def \bjtitle#1{#1}\fi
\ifx \bvolume  \undefined \def \bvolume#1{\textbf{#1}}\fi
\ifx \byear  \undefined \def \byear#1{#1}\fi
\ifx \bissue  \undefined \def \bissue#1{#1}\fi
\ifx \bfpage  \undefined \def \bfpage#1{#1}\fi
\ifx \blpage  \undefined \def \blpage #1{#1}\fi
\ifx \burl  \undefined \def \burl#1{\textsf{#1}}\fi
\ifx \doiurl  \undefined \def \doiurl#1{\url{https://doi.org/#1}}\fi
\ifx \betal  \undefined \def \betal{\textit{et al.}}\fi
\ifx \binstitute  \undefined \def \binstitute#1{#1}\fi
\ifx \binstitutionaled  \undefined \def \binstitutionaled#1{#1}\fi
\ifx \bctitle  \undefined \def \bctitle#1{#1}\fi
\ifx \beditor  \undefined \def \beditor#1{#1}\fi
\ifx \bpublisher  \undefined \def \bpublisher#1{#1}\fi
\ifx \bbtitle  \undefined \def \bbtitle#1{#1}\fi
\ifx \bedition  \undefined \def \bedition#1{#1}\fi
\ifx \bseriesno  \undefined \def \bseriesno#1{#1}\fi
\ifx \blocation  \undefined \def \blocation#1{#1}\fi
\ifx \bsertitle  \undefined \def \bsertitle#1{#1}\fi
\ifx \bsnm \undefined \def \bsnm#1{#1}\fi
\ifx \bsuffix \undefined \def \bsuffix#1{#1}\fi
\ifx \bparticle \undefined \def \bparticle#1{#1}\fi
\ifx \barticle \undefined \def \barticle#1{#1}\fi
\bibcommenthead
\ifx \bconfdate \undefined \def \bconfdate #1{#1}\fi
\ifx \botherref \undefined \def \botherref #1{#1}\fi
\ifx \url \undefined \def \url#1{\textsf{#1}}\fi
\ifx \bchapter \undefined \def \bchapter#1{#1}\fi
\ifx \bbook \undefined \def \bbook#1{#1}\fi
\ifx \bcomment \undefined \def \bcomment#1{#1}\fi
\ifx \oauthor \undefined \def \oauthor#1{#1}\fi
\ifx \citeauthoryear \undefined \def \citeauthoryear#1{#1}\fi
\ifx \endbibitem  \undefined \def \endbibitem {}\fi
\ifx \bconflocation  \undefined \def \bconflocation#1{#1}\fi
\ifx \arxivurl  \undefined \def \arxivurl#1{\textsf{#1}}\fi
\csname PreBibitemsHook\endcsname

\bibitem[\protect\citeauthoryear{Cai et~al.}{2021}]{cai2021physics}
\begin{barticle}
\bauthor{\bsnm{Cai}, \binits{S.}},
\bauthor{\bsnm{Wang}, \binits{Z.}},
\bauthor{\bsnm{Wang}, \binits{S.}},
\bauthor{\bsnm{Perdikaris}, \binits{P.}},
\bauthor{\bsnm{Karniadakis}, \binits{G.E.}}:
\batitle{Physics-informed neural networks for heat transfer problems}.
\bjtitle{Journal of Heat Transfer}
\bvolume{143}(\bissue{6}),
\bfpage{060801}
(\byear{2021})
\end{barticle}
\endbibitem

\bibitem[\protect\citeauthoryear{Wu et~al.}{2019}]{wu2019predicting}
\begin{barticle}
\bauthor{\bsnm{Wu}, \binits{Y.-J.}},
\bauthor{\bsnm{Fang}, \binits{L.}},
\bauthor{\bsnm{Xu}, \binits{Y.}}:
\batitle{Predicting interfacial thermal resistance by machine learning}.
\bjtitle{npj Computational Materials}
\bvolume{5}(\bissue{1}),
\bfpage{56}
(\byear{2019})
\end{barticle}
\endbibitem

\bibitem[\protect\citeauthoryear{He et~al.}{2024}]{he2024multi}
\begin{barticle}
\bauthor{\bsnm{He}, \binits{W.}},
\bauthor{\bsnm{Li}, \binits{J.}},
\bauthor{\bsnm{Kong}, \binits{X.}},
\bauthor{\bsnm{Deng}, \binits{L.}}:
\batitle{Multi-level physics informed deep learning for solving partial differential equations in computational structural mechanics}.
\bjtitle{Communications Engineering}
\bvolume{3}(\bissue{1}),
\bfpage{151}
(\byear{2024})
\end{barticle}
\endbibitem

\bibitem[\protect\citeauthoryear{Li et~al.}{2024}]{li2024mechanics}
\begin{barticle}
\bauthor{\bsnm{Li}, \binits{X.}},
\bauthor{\bsnm{Bolandi}, \binits{H.}},
\bauthor{\bsnm{Masmoudi}, \binits{M.}},
\bauthor{\bsnm{Salem}, \binits{T.}},
\bauthor{\bsnm{Jha}, \binits{A.}},
\bauthor{\bsnm{Lajnef}, \binits{N.}},
\bauthor{\bsnm{Boddeti}, \binits{V.N.}}:
\batitle{Mechanics-informed autoencoder enables automated detection and localization of unforeseen structural damage}.
\bjtitle{Nature Communications}
\bvolume{15}(\bissue{1}),
\bfpage{9229}
(\byear{2024})
\end{barticle}
\endbibitem

\bibitem[\protect\citeauthoryear{Font et~al.}{2025}]{font2025deep}
\begin{barticle}
\bauthor{\bsnm{Font}, \binits{B.}},
\bauthor{\bsnm{Alc{\'a}ntara-{\'A}vila}, \binits{F.}},
\bauthor{\bsnm{Rabault}, \binits{J.}},
\bauthor{\bsnm{Vinuesa}, \binits{R.}},
\bauthor{\bsnm{Lehmkuhl}, \binits{O.}}:
\batitle{Deep reinforcement learning for active flow control in a turbulent separation bubble}.
\bjtitle{Nature communications}
\bvolume{16}(\bissue{1}),
\bfpage{1422}
(\byear{2025})
\end{barticle}
\endbibitem

\bibitem[\protect\citeauthoryear{Li et~al.}{2024}]{li2024synthetic}
\begin{barticle}
\bauthor{\bsnm{Li}, \binits{T.}},
\bauthor{\bsnm{Biferale}, \binits{L.}},
\bauthor{\bsnm{Bonaccorso}, \binits{F.}},
\bauthor{\bsnm{Scarpolini}, \binits{M.A.}},
\bauthor{\bsnm{Buzzicotti}, \binits{M.}}:
\batitle{Synthetic lagrangian turbulence by generative diffusion models}.
\bjtitle{Nature Machine Intelligence}
\bvolume{6}(\bissue{4}),
\bfpage{393}--\blpage{403}
(\byear{2024})
\end{barticle}
\endbibitem

\bibitem[\protect\citeauthoryear{Zhu et~al.}{2025}]{zhu2025frequency}
\begin{barticle}
\bauthor{\bsnm{Zhu}, \binits{E.}},
\bauthor{\bsnm{Zong}, \binits{Z.}},
\bauthor{\bsnm{Li}, \binits{E.}},
\bauthor{\bsnm{Lu}, \binits{Y.}},
\bauthor{\bsnm{Zhang}, \binits{J.}},
\bauthor{\bsnm{Xie}, \binits{H.}},
\bauthor{\bsnm{Li}, \binits{Y.}},
\bauthor{\bsnm{Yin}, \binits{W.-Y.}},
\bauthor{\bsnm{Wei}, \binits{Z.}}:
\batitle{Frequency transfer and inverse design for metasurface under multi-physics coupling by euler latent dynamic and data-analytical regularizations}.
\bjtitle{Nature Communications}
\bvolume{16}(\bissue{1}),
\bfpage{2251}
(\byear{2025})
\end{barticle}
\endbibitem

\bibitem[\protect\citeauthoryear{Trivedi et~al.}{2019}]{trivedi2019data}
\begin{barticle}
\bauthor{\bsnm{Trivedi}, \binits{R.}},
\bauthor{\bsnm{Su}, \binits{L.}},
\bauthor{\bsnm{Lu}, \binits{J.}},
\bauthor{\bsnm{Schubert}, \binits{M.F.}},
\bauthor{\bsnm{Vuckovic}, \binits{J.}}:
\batitle{Data-driven acceleration of photonic simulations}.
\bjtitle{Scientific reports}
\bvolume{9}(\bissue{1}),
\bfpage{19728}
(\byear{2019})
\end{barticle}
\endbibitem

\bibitem[\protect\citeauthoryear{Lu et~al.}{2019}]{lu2019deeponet}
\begin{botherref}
\oauthor{\bsnm{Lu}, \binits{L.}},
\oauthor{\bsnm{Jin}, \binits{P.}},
\oauthor{\bsnm{Karniadakis}, \binits{G.E.}}:
Deeponet: Learning nonlinear operators for identifying differential equations based on the universal approximation theorem of operators.
arXiv preprint arXiv:1910.03193
(2019)
\end{botherref}
\endbibitem

\bibitem[\protect\citeauthoryear{Hsieh et~al.}{2019}]{hsieh2019learning}
\begin{botherref}
\oauthor{\bsnm{Hsieh}, \binits{J.-T.}},
\oauthor{\bsnm{Zhao}, \binits{S.}},
\oauthor{\bsnm{Eismann}, \binits{S.}},
\oauthor{\bsnm{Mirabella}, \binits{L.}},
\oauthor{\bsnm{Ermon}, \binits{S.}}:
Learning neural pde solvers with convergence guarantees.
arXiv preprint arXiv:1906.01200
(2019)
\end{botherref}
\endbibitem

\bibitem[\protect\citeauthoryear{Kopani{\v{c}}{\'a}kov{\'a} and Karniadakis}{2025}]{kopanivcakova2025deeponet}
\begin{barticle}
\bauthor{\bsnm{Kopani{\v{c}}{\'a}kov{\'a}}, \binits{A.}},
\bauthor{\bsnm{Karniadakis}, \binits{G.E.}}:
\batitle{Deeponet based preconditioning strategies for solving parametric linear systems of equations}.
\bjtitle{SIAM Journal on Scientific Computing}
\bvolume{47}(\bissue{1}),
\bfpage{151}--\blpage{181}
(\byear{2025})
\end{barticle}
\endbibitem

\bibitem[\protect\citeauthoryear{Luo et~al.}{2024}]{luo2024neural}
\begin{barticle}
\bauthor{\bsnm{Luo}, \binits{J.}},
\bauthor{\bsnm{Wang}, \binits{J.}},
\bauthor{\bsnm{Wang}, \binits{H.}},
\bauthor{\bsnm{Geng}, \binits{Z.}},
\bauthor{\bsnm{Chen}, \binits{H.}},
\bauthor{\bsnm{Kuang}, \binits{Y.}}, \betal:
\batitle{Neural krylov iteration for accelerating linear system solving}.
\bjtitle{Advances in Neural Information Processing Systems}
\bvolume{37},
\bfpage{128636}--\blpage{128667}
(\byear{2024})
\end{barticle}
\endbibitem

\bibitem[\protect\citeauthoryear{Trifonov et~al.}{2024}]{trifonov2024learning}
\begin{botherref}
\oauthor{\bsnm{Trifonov}, \binits{V.}},
\oauthor{\bsnm{Rudikov}, \binits{A.}},
\oauthor{\bsnm{Iliev}, \binits{O.}},
\oauthor{\bsnm{Laevsky}, \binits{Y.M.}},
\oauthor{\bsnm{Oseledets}, \binits{I.}},
\oauthor{\bsnm{Muravleva}, \binits{E.}}:
Learning from linear algebra: A graph neural network approach to preconditioner design for conjugate gradient solvers.
arXiv preprint arXiv:2405.15557
(2024)
\end{botherref}
\endbibitem

\bibitem[\protect\citeauthoryear{Raissi et~al.}{2019}]{raissi2019physics}
\begin{barticle}
\bauthor{\bsnm{Raissi}, \binits{M.}},
\bauthor{\bsnm{Perdikaris}, \binits{P.}},
\bauthor{\bsnm{Karniadakis}, \binits{G.E.}}:
\batitle{Physics-informed neural networks: A deep learning framework for solving forward and inverse problems involving nonlinear partial differential equations}.
\bjtitle{Journal of Computational physics}
\bvolume{378},
\bfpage{686}--\blpage{707}
(\byear{2019})
\end{barticle}
\endbibitem

\bibitem[\protect\citeauthoryear{Li et~al.}{2020}]{li2020fourier}
\begin{botherref}
\oauthor{\bsnm{Li}, \binits{Z.}},
\oauthor{\bsnm{Kovachki}, \binits{N.}},
\oauthor{\bsnm{Azizzadenesheli}, \binits{K.}},
\oauthor{\bsnm{Liu}, \binits{B.}},
\oauthor{\bsnm{Bhattacharya}, \binits{K.}},
\oauthor{\bsnm{Stuart}, \binits{A.}},
\oauthor{\bsnm{Anandkumar}, \binits{A.}}:
Fourier neural operator for parametric partial differential equations.
arXiv preprint arXiv:2010.08895
(2020)
\end{botherref}
\endbibitem

\bibitem[\protect\citeauthoryear{Cao et~al.}{2024}]{cao2024laplace}
\begin{barticle}
\bauthor{\bsnm{Cao}, \binits{Q.}},
\bauthor{\bsnm{Goswami}, \binits{S.}},
\bauthor{\bsnm{Karniadakis}, \binits{G.E.}}:
\batitle{Laplace neural operator for solving differential equations}.
\bjtitle{Nature Machine Intelligence}
\bvolume{6}(\bissue{6}),
\bfpage{631}--\blpage{640}
(\byear{2024})
\end{barticle}
\endbibitem

\bibitem[\protect\citeauthoryear{Sanchez-Gonzalez et~al.}{2020}]{sanchez2020learning}
\begin{bchapter}
\bauthor{\bsnm{Sanchez-Gonzalez}, \binits{A.}},
\bauthor{\bsnm{Godwin}, \binits{J.}},
\bauthor{\bsnm{Pfaff}, \binits{T.}},
\bauthor{\bsnm{Ying}, \binits{R.}},
\bauthor{\bsnm{Leskovec}, \binits{J.}},
\bauthor{\bsnm{Battaglia}, \binits{P.}}:
\bctitle{Learning to simulate complex physics with graph networks}.
In: \bbtitle{International Conference on Machine Learning},
pp. \bfpage{8459}--\blpage{8468}
(\byear{2020}).
\bcomment{PMLR}
\end{bchapter}
\endbibitem

\bibitem[\protect\citeauthoryear{Bryutkin et~al.}{2024}]{bryutkin2024hamlet}
\begin{botherref}
\oauthor{\bsnm{Bryutkin}, \binits{A.}},
\oauthor{\bsnm{Huang}, \binits{J.}},
\oauthor{\bsnm{Deng}, \binits{Z.}},
\oauthor{\bsnm{Yang}, \binits{G.}},
\oauthor{\bsnm{Sch{\"o}nlieb}, \binits{C.-B.}},
\oauthor{\bsnm{Aviles-Rivero}, \binits{A.}}:
Hamlet: Graph transformer neural operator for partial differential equations.
arXiv preprint arXiv:2402.03541
(2024)
\end{botherref}
\endbibitem

\bibitem[\protect\citeauthoryear{Dong et~al.}{2024}]{dong2024pinn}
\begin{botherref}
\oauthor{\bsnm{Dong}, \binits{D.}},
\oauthor{\bsnm{Suo}, \binits{W.}},
\oauthor{\bsnm{Kou}, \binits{J.}},
\oauthor{\bsnm{Zhang}, \binits{W.}}:
Pinn-mg: A multigrid-inspired hybrid framework combining iterative method and physics-informed neural networks.
arXiv preprint arXiv:2410.05744
(2024)
\end{botherref}
\endbibitem

\bibitem[\protect\citeauthoryear{Zhang et~al.}{2024}]{zhang2024blending}
\begin{botherref}
\oauthor{\bsnm{Zhang}, \binits{E.}},
\oauthor{\bsnm{Kahana}, \binits{A.}},
\oauthor{\bsnm{Kopani{\v{c}}{\'a}kov{\'a}}, \binits{A.}},
\oauthor{\bsnm{Turkel}, \binits{E.}},
\oauthor{\bsnm{Ranade}, \binits{R.}},
\oauthor{\bsnm{Pathak}, \binits{J.}},
\oauthor{\bsnm{Karniadakis}, \binits{G.E.}}:
Blending neural operators and relaxation methods in pde numerical solvers.
Nature Machine Intelligence,
1--11
(2024)
\end{botherref}
\endbibitem

\bibitem[\protect\citeauthoryear{Jagtap and Karniadakis}{2020}]{jagtap2020extended}
\begin{botherref}
\oauthor{\bsnm{Jagtap}, \binits{A.D.}},
\oauthor{\bsnm{Karniadakis}, \binits{G.E.}}:
Extended physics-informed neural networks (xpinns): A generalized space-time domain decomposition based deep learning framework for nonlinear partial differential equations.
Communications in Computational Physics
\textbf{28}(5)
(2020)
\end{botherref}
\endbibitem

\bibitem[\protect\citeauthoryear{Basir and Senocak}{2023}]{basir2023generalized}
\begin{botherref}
\oauthor{\bsnm{Basir}, \binits{S.}},
\oauthor{\bsnm{Senocak}, \binits{I.}}:
A generalized schwarz-type non-overlapping domain decomposition method using physics-constrained neural networks.
arXiv preprint arXiv:2307.12435
(2023)
\end{botherref}
\endbibitem

\bibitem[\protect\citeauthoryear{Dolean et~al.}{2024}]{dolean2024multilevel}
\begin{barticle}
\bauthor{\bsnm{Dolean}, \binits{V.}},
\bauthor{\bsnm{Heinlein}, \binits{A.}},
\bauthor{\bsnm{Mishra}, \binits{S.}},
\bauthor{\bsnm{Moseley}, \binits{B.}}:
\batitle{Multilevel domain decomposition-based architectures for physics-informed neural networks}.
\bjtitle{Computer Methods in Applied Mechanics and Engineering}
\bvolume{429},
\bfpage{117116}
(\byear{2024})
\end{barticle}
\endbibitem

\bibitem[\protect\citeauthoryear{Mao et~al.}{2024}]{pmlr-v235-mao24b}
\begin{bchapter}
\bauthor{\bsnm{Mao}, \binits{C.}},
\bauthor{\bsnm{Lupoiu}, \binits{R.}},
\bauthor{\bsnm{Dai}, \binits{T.}},
\bauthor{\bsnm{Chen}, \binits{M.}},
\bauthor{\bsnm{Fan}, \binits{J.}}:
\bctitle{Towards general neural surrogate solvers with specialized neural accelerators}.
In: \beditor{\bsnm{Salakhutdinov}, \binits{R.}},
\beditor{\bsnm{Kolter}, \binits{Z.}},
\beditor{\bsnm{Heller}, \binits{K.}},
\beditor{\bsnm{Weller}, \binits{A.}},
\beditor{\bsnm{Oliver}, \binits{N.}},
\beditor{\bsnm{Scarlett}, \binits{J.}},
\beditor{\bsnm{Berkenkamp}, \binits{F.}} (eds.)
\bbtitle{Proceedings of the 41st International Conference on Machine Learning}.
\bsertitle{Proceedings of Machine Learning Research},
vol. \bseriesno{235},
pp. \bfpage{34693}--\blpage{34711}.
\bpublisher{PMLR}, \blocation{???}
(\byear{2024}).
\burl{https://proceedings.mlr.press/v235/mao24b.html}
\end{bchapter}
\endbibitem

\bibitem[\protect\citeauthoryear{Luz et~al.}{2020}]{luz2020learning}
\begin{bchapter}
\bauthor{\bsnm{Luz}, \binits{I.}},
\bauthor{\bsnm{Galun}, \binits{M.}},
\bauthor{\bsnm{Maron}, \binits{H.}},
\bauthor{\bsnm{Basri}, \binits{R.}},
\bauthor{\bsnm{Yavneh}, \binits{I.}}:
\bctitle{Learning algebraic multigrid using graph neural networks}.
In: \bbtitle{International Conference on Machine Learning},
pp. \bfpage{6489}--\blpage{6499}
(\byear{2020}).
\bcomment{PMLR}
\end{bchapter}
\endbibitem

\bibitem[\protect\citeauthoryear{Lerer et~al.}{2024}]{lerer2024multigrid}
\begin{barticle}
\bauthor{\bsnm{Lerer}, \binits{B.}},
\bauthor{\bsnm{Ben-Yair}, \binits{I.}},
\bauthor{\bsnm{Treister}, \binits{E.}}:
\batitle{Multigrid-augmented deep learning preconditioners for the helmholtz equation using compact implicit layers}.
\bjtitle{SIAM Journal on Scientific Computing}
\bvolume{46}(\bissue{5}),
\bfpage{123}--\blpage{144}
(\byear{2024})
\end{barticle}
\endbibitem

\bibitem[\protect\citeauthoryear{Heinlein et~al.}{2021}]{heinlein2021combining}
\begin{barticle}
\bauthor{\bsnm{Heinlein}, \binits{A.}},
\bauthor{\bsnm{Klawonn}, \binits{A.}},
\bauthor{\bsnm{Lanser}, \binits{M.}},
\bauthor{\bsnm{Weber}, \binits{J.}}:
\batitle{Combining machine learning and adaptive coarse spaces---a hybrid approach for robust feti-dp methods in three dimensions}.
\bjtitle{SIAM Journal on Scientific Computing}
\bvolume{43}(\bissue{5}),
\bfpage{816}--\blpage{838}
(\byear{2021})
\end{barticle}
\endbibitem

\bibitem[\protect\citeauthoryear{Heinlein et~al.}{2019}]{heinlein2019machine}
\begin{barticle}
\bauthor{\bsnm{Heinlein}, \binits{A.}},
\bauthor{\bsnm{Klawonn}, \binits{A.}},
\bauthor{\bsnm{Lanser}, \binits{M.}},
\bauthor{\bsnm{Weber}, \binits{J.}}:
\batitle{Machine learning in adaptive domain decomposition methods---predicting the geometric location of constraints}.
\bjtitle{SIAM Journal on Scientific Computing}
\bvolume{41}(\bissue{6}),
\bfpage{3887}--\blpage{3912}
(\byear{2019})
\end{barticle}
\endbibitem

\bibitem[\protect\citeauthoryear{Heinlein et~al.}{2023}]{heinlein2023predicting}
\begin{bchapter}
\bauthor{\bsnm{Heinlein}, \binits{A.}},
\bauthor{\bsnm{Klawonn}, \binits{A.}},
\bauthor{\bsnm{Lanser}, \binits{M.}},
\bauthor{\bsnm{Weber}, \binits{J.}}:
\bctitle{Predicting the geometric location of critical edges in adaptive gdsw overlapping domain decomposition methods using deep learning}.
In: \bbtitle{Domain Decomposition Methods in Science and Engineering XXVI},
pp. \bfpage{307}--\blpage{315}.
\bpublisher{Springer}, \blocation{???}
(\byear{2023})
\end{bchapter}
\endbibitem

\bibitem[\protect\citeauthoryear{Azulay and Treister}{2022}]{azulay2022multigrid}
\begin{barticle}
\bauthor{\bsnm{Azulay}, \binits{Y.}},
\bauthor{\bsnm{Treister}, \binits{E.}}:
\batitle{Multigrid-augmented deep learning preconditioners for the helmholtz equation}.
\bjtitle{SIAM Journal on Scientific Computing}
\bvolume{45}(\bissue{3}),
\bfpage{127}--\blpage{151}
(\byear{2022})
\end{barticle}
\endbibitem

\bibitem[\protect\citeauthoryear{Cui et~al.}{2024}]{cui2024neural}
\begin{botherref}
\oauthor{\bsnm{Cui}, \binits{C.}},
\oauthor{\bsnm{Jiang}, \binits{K.}},
\oauthor{\bsnm{Shu}, \binits{S.}}:
A neural multigrid solver for helmholtz equations with high wavenumber and heterogeneous media.
arXiv preprint arXiv:2404.02493
(2024)
\end{botherref}
\endbibitem

\bibitem[\protect\citeauthoryear{Xie et~al.}{2025}]{xiemgcfnn}
\begin{bchapter}
\bauthor{\bsnm{Xie}, \binits{Y.}},
\bauthor{\bsnm{Lv}, \binits{M.}},
\bauthor{\bsnm{Zhang}, \binits{C.-S.}}:
\bctitle{Mgcfnn: A neural multigrid solver with novel fourier neural network for high wave number helmholtz equations}.
In: \bbtitle{The Thirteenth International Conference on Learning Representations}
(\byear{2025})
\end{bchapter}
\endbibitem

\bibitem[\protect\citeauthoryear{Knoke et~al.}{2023}]{knoke2023domain}
\begin{botherref}
\oauthor{\bsnm{Knoke}, \binits{T.}},
\oauthor{\bsnm{Kinnewig}, \binits{S.}},
\oauthor{\bsnm{Beuchler}, \binits{S.}},
\oauthor{\bsnm{Demircan}, \binits{A.}},
\oauthor{\bsnm{Morgner}, \binits{U.}},
\oauthor{\bsnm{Wick}, \binits{T.}}:
Domain decomposition with neural network interface approximations for time-harmonic maxwell's equations with different wave numbers.
arXiv preprint arXiv:2303.02590
(2023)
\end{botherref}
\endbibitem

\bibitem[\protect\citeauthoryear{Piao et~al.}{2024}]{piao2024domain}
\begin{barticle}
\bauthor{\bsnm{Piao}, \binits{S.}},
\bauthor{\bsnm{Gu}, \binits{H.}},
\bauthor{\bsnm{Wang}, \binits{A.}},
\bauthor{\bsnm{Qin}, \binits{P.}}:
\batitle{A domain-adaptive physics-informed neural network for inverse problems of maxwell's equations in heterogeneous media}.
\bjtitle{IEEE Antennas and Wireless Propagation Letters}
\bvolume{23}(\bissue{10}),
\bfpage{2905}--\blpage{2909}
(\byear{2024})
\end{barticle}
\endbibitem

\bibitem[\protect\citeauthoryear{Dalcin et~al.}{2011}]{dalcin2011parallel}
\begin{barticle}
\bauthor{\bsnm{Dalcin}, \binits{L.D.}},
\bauthor{\bsnm{Paz}, \binits{R.R.}},
\bauthor{\bsnm{Kler}, \binits{P.A.}},
\bauthor{\bsnm{Cosimo}, \binits{A.}}:
\batitle{Parallel distributed computing using python}.
\bjtitle{Advances in Water Resources}
\bvolume{34}(\bissue{9}),
\bfpage{1124}--\blpage{1139}
(\byear{2011})
\end{barticle}
\endbibitem

\bibitem[\protect\citeauthoryear{Dolean et~al.}{2015}]{dolean2015introduction}
\begin{bbook}
\bauthor{\bsnm{Dolean}, \binits{V.}},
\bauthor{\bsnm{Jolivet}, \binits{P.}},
\bauthor{\bsnm{Nataf}, \binits{F.}}:
\bbtitle{An Introduction to Domain Decomposition Methods: Algorithms, Theory, and Parallel Implementation}.
\bpublisher{SIAM}, \blocation{???}
(\byear{2015})
\end{bbook}
\endbibitem

\bibitem[\protect\citeauthoryear{Nataf et~al.}{2010}]{nataf2010two}
\begin{barticle}
\bauthor{\bsnm{Nataf}, \binits{F.}},
\bauthor{\bsnm{Xiang}, \binits{H.}},
\bauthor{\bsnm{Dolean}, \binits{V.}}:
\batitle{A two level domain decomposition preconditioner based on local dirichlet-to-neumann maps}.
\bjtitle{Comptes Rendus. Math{\'e}matique}
\bvolume{348}(\bissue{21-22}),
\bfpage{1163}--\blpage{1167}
(\byear{2010})
\end{barticle}
\endbibitem

\bibitem[\protect\citeauthoryear{Haferssas et~al.}{2015}]{haferssas2015robust}
\begin{barticle}
\bauthor{\bsnm{Haferssas}, \binits{R.}},
\bauthor{\bsnm{Jolivet}, \binits{P.}},
\bauthor{\bsnm{Nataf}, \binits{F.}}:
\batitle{A robust coarse space for optimized schwarz methods: Soras-geneo-2}.
\bjtitle{Comptes Rendus Mathematique}
\bvolume{353}(\bissue{10}),
\bfpage{959}--\blpage{963}
(\byear{2015})
\end{barticle}
\endbibitem

\bibitem[\protect\citeauthoryear{Farhat et~al.}{2000}]{farhat2000two}
\begin{barticle}
\bauthor{\bsnm{Farhat}, \binits{C.}},
\bauthor{\bsnm{Macedo}, \binits{A.}},
\bauthor{\bsnm{Lesoinne}, \binits{M.}}:
\batitle{A two-level domain decomposition method for the iterative solution of high frequency exterior helmholtz problems}.
\bjtitle{Numerische Mathematik}
\bvolume{85}(\bissue{2}),
\bfpage{283}--\blpage{308}
(\byear{2000})
\end{barticle}
\endbibitem

\bibitem[\protect\citeauthoryear{Schenk et~al.}{2001}]{schenk2001pardiso}
\begin{barticle}
\bauthor{\bsnm{Schenk}, \binits{O.}},
\bauthor{\bsnm{G{\"a}rtner}, \binits{K.}},
\bauthor{\bsnm{Fichtner}, \binits{W.}},
\bauthor{\bsnm{Stricker}, \binits{A.}}:
\batitle{Pardiso: a high-performance serial and parallel sparse linear solver in semiconductor device simulation}.
\bjtitle{Future Generation Computer Systems}
\bvolume{18}(\bissue{1}),
\bfpage{69}--\blpage{78}
(\byear{2001})
\end{barticle}
\endbibitem

\bibitem[\protect\citeauthoryear{Hughes et~al.}{2019}]{hughes2019forward}
\begin{barticle}
\bauthor{\bsnm{Hughes}, \binits{T.W.}},
\bauthor{\bsnm{Williamson}, \binits{I.A.}},
\bauthor{\bsnm{Minkov}, \binits{M.}},
\bauthor{\bsnm{Fan}, \binits{S.}}:
\batitle{Forward-mode differentiation of maxwell’s equations}.
\bjtitle{ACS Photonics}
\bvolume{6}(\bissue{11}),
\bfpage{3010}--\blpage{3016}
(\byear{2019})
\end{barticle}
\endbibitem

\end{thebibliography}

\end{document}